\documentclass{article}

     \PassOptionsToPackage{numbers, compress}{natbib}
\usepackage[final]{neurips_2022}

\usepackage[utf8]{inputenc} % allow utf-8 input
\usepackage[T1]{fontenc}    % use 8-bit T1 fonts
\usepackage{hyperref}       % hyperlinks
\usepackage{url}            % simple URL typesetting
\usepackage{booktabs}       % professional-quality tables
\usepackage{amsfonts}       % blackboard math symbols
\usepackage{nicefrac}       % compact symbols for 1/2, etc.
\usepackage{microtype}      % microtypography
\usepackage{xcolor}         % colors
\usepackage[pdftex]{graphicx}

\usepackage{csquotes}

\usepackage{etoolbox,xspace}
\providetoggle{anom}
\settoggle{anom}{false}

\newcommand{\dataset}[1]{\textsc{#1}\xspace}
\newcommand{\eurocrops}{\iftoggle{anom}{\dataset{AnomDataset}}{{EuroCrops}}\xspace}

\title{Challenges and Opportunities of \\ Large Transnational Datasets: \\ A Case Study on European Administrative Crop Data}

\author{%
  Maja Schneider \\
  Technical University of Munich (TUM)\\
  TUM School of Engineering and Design\\
  80333 Munich, Germany \\
  \texttt{maja.schneider@tum.de} \\
  \And
  Christian Marchington \\
  London, UK \\
  \texttt{christian@marchington.dev} \\
  \And
  Marco K{\"o}rner \\
  Technical University of Munich (TUM)\\
  TUM School of Engineering and Design\\
  80333 Munich, Germany \\
  \texttt{marco.koerner@tum.de} \\
}

\begin{document}

\maketitle

\begin{abstract}
    Expansive, informative datasets are vital in providing foundations and possibilities for scientific research and development across many fields of study.
    Assembly of grand datasets, however, frequently poses difficulty for the author and stakeholders alike, with a variety of considerations required throughout the collaboration efforts and development lifecycle.
    In this work, we discuss and analyse the challenges and opportunities we faced throughout the creation of a transnational, European agricultural dataset containing reference labels of cultivated crops.
    Together, this forms a succinct framework of important elements one should consider when forging a dataset of their own.
\end{abstract}

\section{Introduction}

With the progression of member states of the \emph{European Union (EU)} to continually and openly publish administrative data, the opportunity to conduct research previously limited to a local-scale has advanced massively with the emergence of expansive, collaborative, and multi-national datasets.
The \eurocrops initiative%
\iftoggle{anom}{\footnote{Dataset name and references hidden for blind review.}}{\cite{schneider2021TEC, schneider2022eurocrops21, schneider2021EPE}}
harnessed the continually improving accessibility to \emph{Common agriculture policy (CAP)} \cite{ec2022cap} data, demonstrating the feasibility of a transnational dataset and building out a framework of considerations one must account for when developing projects at a similar spatial coverage.
While there are clear advantages of a large and diverse dataset for all breadths of research, the number of challenges imposed by administrations’ legacy, and often manual, systems continues to make gathering data at this scale an ever-laborious task.
\Citet{marini2020BDE}, for instance, face these obstacles when they harmonised several sub-databases containing real estate information in order to analyse the social and economic changes within Europe. %\cite{marini2020BDE}.
More generally, \citet{connelly2016RAD} analysed the issues with administrative social science data while stressing the importance of its use. %\cite{connelly2016RAD}.

In this paper, we distilled the challenges we faced while working on \eurocrops into a framework consisting of six distinct categories, which will be further explored in the sections following.
The goal of this framework is to provide researchers and authorities with guidelines when approaching transnational, country-dependent, and collaborative open data projects.

While highlighting the challenges involved, we also want to encourage the community to take part in the development of large pan-European datasets by showcasing the advantages and benefits they are able to provide for a multitude of stakeholders.

\section{The Study Data}

%\begin{figure}[t]
%    \centering
%    \includegraphics[width=0.8\linewidth]{map.pdf}
%    \caption{The \eurocrops dataset now includes data from 16 member states of the \emph{European Union (EU)}. The countries coloured in green provide data and helped with the identification of the six biggest challenges faced when compiling collaborative, transnational datasets, which are listed on the left. At the time of the first \eurocrops release, data from the red countries was not available yet.}
%    \label{fig:map}
%\end{figure}

\begin{figure}[t]
    \centering
    \includegraphics[width=0.8\linewidth]{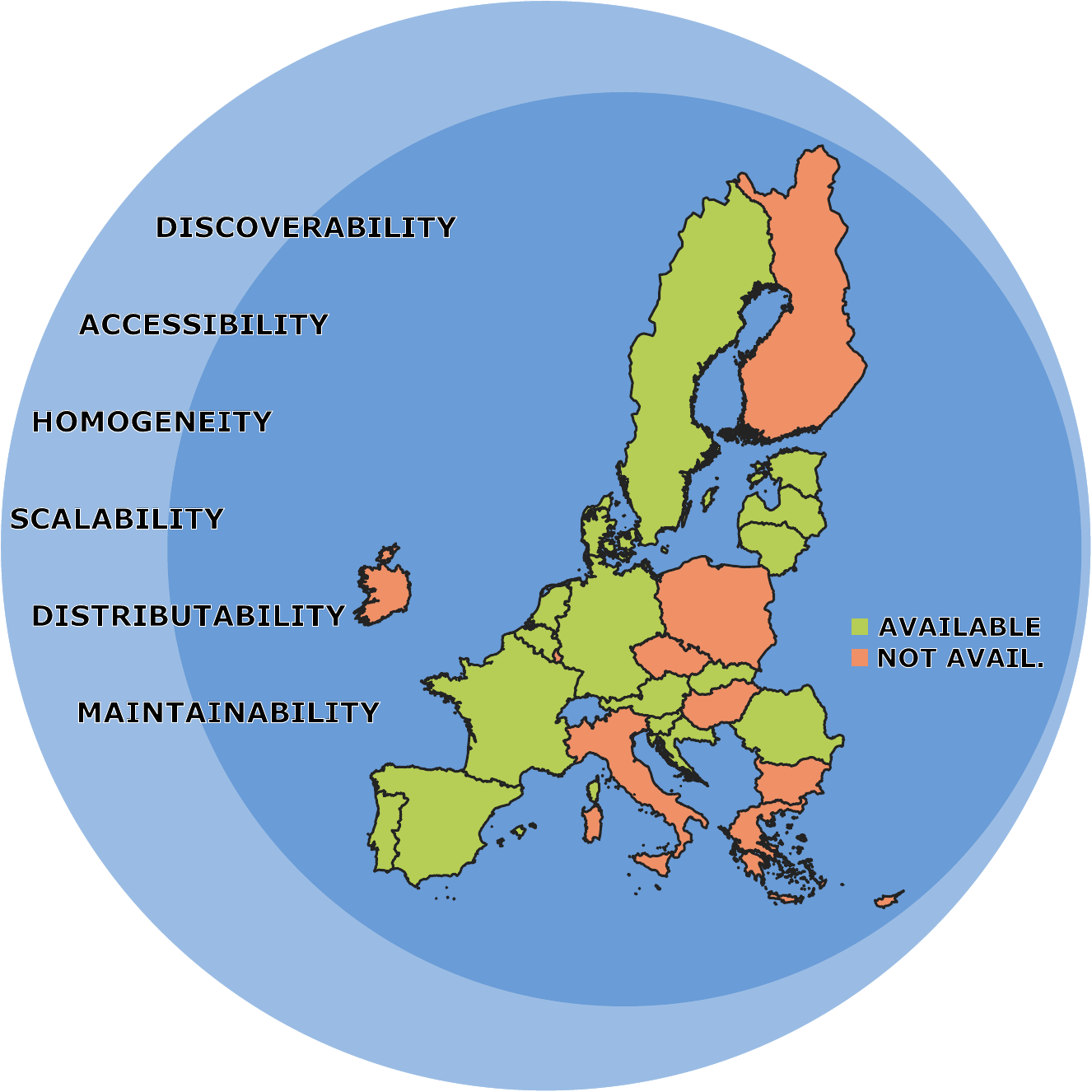}
    \caption{The \eurocrops dataset now includes data from 16 member states of the \emph{European Union (EU)}. The countries coloured in green provide data and helped with the identification of the six biggest challenges faced when compiling collaborative, transnational datasets, which are listed on the left. At the time of the first \eurocrops release, data from the red countries was not available yet.}
    \label{fig:map}
\end{figure}

\eurocrops can be considered to be the first large-scale, pan-European dataset for cultivated crops, following the approaches of the LUCAS \cite{andrimont2020harmonised} and BreizhCrops \cite{russwurm2019breizhcrops} initiatives, combining and harmonising georeferenced agricultural parcel data with the corresponding crop.
While originally motivated as a use-case for applying machine learning methods to Earth observation data, more specifically Copernicus Sentinel-2 satellite imagery, \eurocrops has now become an impactful project spanning across a number of disciplines, including biodiversity and agriculture.
The data itself was obtained through the countries’ agricultural ministries or paying agencies and contains the farmers' self-declarations, which are collected within the subsidy control of the \emph{Common agriculture policy (CAP)}.
At the time of this study's publication, 16 member states of the \emph{EU} have contributed data, highlighted in green in Figure \ref{fig:map}, which we collected over a span of 1.5 years.

As part of the collection effort, a new \emph{Hierarchical Crop and Agriculture Taxonomy (HCAT)} \iftoggle{anom}{}{\cite{schneider2022hcat}} was designed to both accommodate the high crop diversity and harmonise country-unique cultivation schemas and languages across Europe. 
The harmonised \eurocrops vector data is centrally and publicly available via the project’s GitHub repository%
\footnote{\iftoggle{anom}{URL hidden for blind review.}{\url{https://github.com/maja601/EuroCrops}}.}, 
on Zenodo%
\footnote{\iftoggle{anom}{URL hidden for blind review.}{\url{https://doi.org/10.5281/zenodo.6866846}}.}, 
and will be distributed over commonly utilised platforms, such as Google Earth Engine%
\footnote{\url{https://earthengine.google.com/}.}, 
GeoDB%
\footnote{\url{https://eurodatacube.com/marketplace/services/edc_geodb}.}, 
CODE-DE%
\footnote{\url{https://code-de.org/en/}.},
and EO-Lab%
\footnote{\url{https://eo-lab.org/en/}.}.

%\begin{figure}[t]
%    \centering
%    \includegraphics[width=0.8\linewidth]{map.pdf}
%    \caption{The \eurocrops dataset now includes data from 16 member states of the \emph{European Union (EU)}. The countries coloured in green provide data and helped with the identification of the six biggest challenges faced when compiling collaborative, transnational datasets, which are listed on the left. At the time of the first \eurocrops release, data from the red countries was not available yet.}
%    \label{fig:map}
%\end{figure}

\section{Challenges}

While compiling the \eurocrops dataset, we identified the following six challenges, forming the foundation of the aforementioned framework.

\subsection{Discoverability}

One would expect that \enquote{open} data would indicate the simplicity of locating the hosted data and determining the type and quantity of the available data; we found this was often not the case.
As countries frequently choose not to use central data distribution points but instead their own national platforms, the method of actually finding these self-hosted repositories still requires incredibly manual investigatory work.
Similar data is often hosted by different types of responsible agencies and authorities across Europe, raising the barrier when attempting to discover the available sources.

Even in the scenarios where the correct authority contact and platform is determined, one still encounters language barriers when navigating local websites, as they are often only hosted in their native language.

\subsection{Accessibility}

The ever-growing concerns surrounding security and storage of personal data continues to motivate authorities and platforms to restrict direct access to the data itself.
This is especially true for data that can be linked back to individuals, notably farmers in the case of \eurocrops, and provides authorities the justification to obscure themselves behind GDPR walls, even if there are legally-just reasons for public access.

Unfortunately, the benefits of taking the step to openly release the data are not justifiably strong enough for the responsible parties to undertake such an effort.
From the outside, there is no observably clear argument against working on the publication, and we can only hypothesise that it might be connected to lack of digitalisation efforts, internal politics, and availability of resources to appropriately prepare the data.
Where resource has been found and appropriately applied, it has sometimes resulted in platforms necessitating user registration before being able to gain access, further adding to the manual effort.

\subsection{Homogeneity}

While lightly touched on previously, an important consideration is the variability in the data content released across countries.
Regarding cultivated vegetation, each country groups their crop data into unique taxonomies and schemes, which further vary between legacy and modern formats.
This includes mixtures of scientific plant nomenclature and terminology in national languages, but also the depth of granularity, ranging from precise biological species to the more generic plant families.

Not only does the data content vary across countries within the EU, but so do the data formats, most notably in fields where severally accepted formats compete within the community.
For example, we observed that, despite being proprietary, Esri shape files still contribute to a large proportion of the geospatial data pool, amongst more modern formats, such as geopackage or webmapservice.
The necessity to handle multiple, conflicting formats when pulling the data from respective authorities further impedes research and development at a transnational scale.

\subsection{Scalability}

In the cases where individual country-dependent datasets are fetched, harmonised, and combined into an aggregated collection, the collection itself is still encapsulated within the pool of available national datasets.
The effort of including new sources then requires restarting this harmonisation process from the beginning, to ensure the content of the additional data complies with the existing collection, while the existing structures have to be altered to accommodate the new dataset.
This is not only true when new countries are added to the transnational collection, but also when ministries or authorities of a certain country migrate from legacy data platforms, or update content structure or format.
These can happen irregularly over the course of years and therefore require manual monitoring and intervention.

\subsection{Distributability}

When releasing or updating the complete dataset, there is the problem of ensuring that all relevant stakeholders receive notification and central access to the data.
Under current circumstances, this typically involves reaching out to each member state individually.
Even though European data repositories exist, these currently lack the functionality to act as a distribution hub, where all information is brought together and subsequently used as an exchange platform for all concerned parties: it is still required to actively engage authorities to ensure they receive the latest updates.

\subsection{Maintainability}

With the increasing size and number of interactions with a dataset, progressively entangled, moving parts begin to co-exist and need to be continuously maintained.
In the case of \eurocrops, a number of country-owned variables will undergo constant development, including: changes to crop policies; infrastructure updates, such as to platforms and authority websites; and standards.
Each of these provides some underlying structure or metadata to the dataset itself, and so require strict manual verification to ensure a high dataset quality is maintained.

When changes to country-owned variables do occur, these are typically carried out without communication, version-control, and tend to be obscured in the background, with error corrections even understated on the authorities' data platforms. 
This makes it extremely challenging to pin down and correct outliers in datasets without significant effort.

\section{Opportunities}

While construction of these large, interwoven datasets presents the array of challenges as discussed above, the opportunities of providing communities with dataset like \eurocrops far outweigh the effort involved.
In the following section, we discuss three of the key target groups and the complementary benefits that the dataset would deliver.

\subsection{For Earth Observation}

As many in the Earth observation community are aware, provision to easily-accessible and complete datasets is often limited and difficult to obtain.
Initiatives to create open-access datasets covering vast swathes of Europe considerably lowers the barrier to entry, providing members of the community with large-scale, manipulable datasets on which they can begin to operate.
These datasets are important, as large-scale Earth observation analysis might require substantially more reference ground data for training machine learning models than one country may be able to provide alone.

Furthermore, Earth observation reference data is often complicated to both understand and work with, being frequently obscured by multiple layers of abstraction and more difficult to find when compared to natural image labels.
By overcoming this hurdle, the domain itself sees further benefits through increased accessibility to outside researchers.

\subsection{For Europe}

A clear benefit of transnational datasets is the weakening of countries’ data sovereignty, which has historically led to encapsulated and incompatible granular datasets and models.

By giving the citizens of the European Union the ability to access and use all available data in a consolidated location and format, research can take a transnational direction, allowing for broader, pan-European problems to be tackled and solutions developed.
This in turn enables far greater potential for both bi- and multilateral research projects and data analysis across borders.

\subsection{For Ministries and Authorities}

When considering ministries and authorities, the benefits of actively participating within transnational datasets take another clear step: data can be accessed, controlled, and maintained within a singular ecosystem; changes can be version-controlled and publicly-stated; and metadata can be standardised.

By providing a unified node that holds all necessary information to download and work with the dataset, authorities are provided with a simple means of interacting with the public and those who request access, as well as establishing a forum for discussion and feedback to be held.

Additionally, with the growth of a singular, established, transnational dataset, there comes the appeal for other countries to participate.
This provides consequent and progressive reward to participants, as they can rely on an existing baseline against which they can fit their data, instead of developing another from the ground up.
Making the entire process far more scalable, both horizontally and vertically, delivers tangible time and effort benefits from the start.

\section{Conclusion}

In this article, we have presented and discussed both the challenges and opportunities that the development of a large-scale, transnational dataset may deliver.

Through the introduction of \eurocrops, we have started to tackle the \textit{discoverability} and \textit{accessibility} barriers by providing one publicly available dataset, which is currently being published on several, well-established and easily discoverable platforms.
The development of \emph{HCAT} expands on this, introducing a sense of \textit{homogeneity} through harmonisation of reference data and allowing for \textit{scalability}, both with new data from existing countries, as well as newly participating countries.

Finally, during the project itself, we managed to build up a community of data providers and data users and form stable connections with a number of authorities and paying agencies across several member states of the European Union.
This way, we hope to actively address the issue of \textit{maintainability} of the dataset and \textit{distributability} of information to all participants of the process, through our now established lines of communication and continually addressing feedback and concerns.

%\begin{ack}
%Use unnumbered first level headings for the acknowledgments. All acknowledgments
%go at the end of the paper before the list of references. Moreover, you are required to declare
%funding (financial activities supporting the submitted work) and competing interests (related financial activities outside the submitted work).
%More information about this disclosure can be found at: \url{https://neurips.cc/Conferences/2022/PaperInformation/FundingDisclosure}.
%
%
%Do {\bf not} include this section in the anonymized submission, only in the final paper. You can use the \texttt{ack} environment provided in the style file to autmoatically hide this section in the anonymized submission.
%\end{ack}

\bibliography{bib}

\end{document}